\documentclass[aps,twocolumn,showpacs]{revtex4}
\usepackage{amsmath}%
\usepackage{amsthm,amsmath,amssymb}
\usepackage{comment}
\usepackage{mathrsfs}
\usepackage{graphicx}
\usepackage{dcolumn}
\usepackage{bm}
\usepackage{color, soul}
\usepackage{color, xcolor}
\usepackage[utf8]{inputenc}
\usepackage[T1]{fontenc}
\usepackage{mathptmx}
\usepackage{multirow}
\usepackage{makecell}
\usepackage{setspace}
\usepackage{footnote}
\usepackage{hyperref}
\hypersetup{hypertex=true,
	colorlinks=true,
	linkcolor=blue,
	anchorcolor=blue,
	citecolor=blue}

\begin{document}
\title{Quantum scattering treatment on the time-domain diffraction of a matter-wave soliton}

\author{Peng Gao$^{1}$}
\author{Jie Liu$^{1}$}\email{jliu@gscaep.ac.cn}

\address{$^1$Graduate School, China Academy of Engineering Physics, Beijing 100193, China}

\begin{abstract}
We study the dynamics of the matter-wave soliton interacting with a vibrating mirror created by an evanescent light and provide a quantum scattering picture for the time-domain diffraction of the matter-wave soliton. Under Kramers-Henneberger (KH) transformation, i.e., in a vibrating coordinate, the vibration of the mirror can be cast to an effective gauge field. We then can exploit Dyson series and the quantum scattering theory to investigate the dynamics of the soliton that moves in the effective gauge field and is reflected by a static mirror. Our analytical theory can quantitatively deduce the locations and the relative weights of the scattered wave packets, which is consistent with our numerical simulations of directly solving a nonlinear Schr\"odinger equation. In particular, for a two-frequency vibrating case, our theory predicts some interesting multi-peak sideband structures in the diffracted matter-wave distributions, which can be resorted to the  resonance of two frequencies. Underlying mechanisms and possible applications are discussed.
\end{abstract}

\maketitle

\section{Introduction}

Manipulation of an atomic wave packet is a topic of great interest  and constantly attracts much attention \cite{Dowling-1996}. An atomic mirror is a feasible way to reflect the wave packets, which can be achieved by a blue-detuning evanescent optical wave made from the total internal reflection of a laser beam in a glass prism \cite{Cook-1982,Balykin-1998}.
More interestingly, when the mirror is vibrating periodically \cite{Henkel-1994}, it offers a scheme for the time-domain diffraction of the atomic wave packets \cite{Steane-1995,Arndt-1995,Colombe-2005}.
Compared with the spatial atomic diffraction using the periodic potential of a crystal surface \cite{Estermann-1930}, a standing wave of light \cite{Moskowitz-1983,Gould-1986}, and fabricated periodic structures \cite{Keith-1988,Carnal-1991}, the time-domain diffraction scheme has an advantage that its diffracted patterns can be readily manipulated by mechanically adjusting the vibrating amplitude and frequency of the glass prism, and therefore arises great interests both theoretically and experimentally \cite{Steane-1995,Arndt-1995,Chen-1996,Felber-1996,Colombe-2005,Xiong-2023}.
Some experiments have been conducted to realize the time-domain diffraction scheme for Cs atoms \cite{Steane-1995,Arndt-1995}, neutrons \cite{Felber-1996}, and Bose-Einstein condensate (BEC) of $^{87}$Rb \cite{Colombe-2005}. In the practical experiments, to avoid the diffusion of a moving wave packet in its free evolution process, a relative larger incident velocity of the matter wave is used and the observed diffraction fringes are in consistence with the semi-classical theory \cite{Henkel-1994,Colombe-2005}. 

A recent work \cite{Xiong-2023} proposes and investigates the matter-wave soliton for the time-domain diffraction scheme, considering that the matter-wave solitons have been widely investigated and generated in diverse BEC systems \cite{Kevrekidis-book,Abdullaev-2005,Frantzeskakis-2010,Kengne-2021,Wu-2002,Zhao-2020,Zhao-2020-2}. The soliton has the property of the high transmission stability, i.e., it can keep its initial wavepacket profile for a long time and not diffuse even with a relative slower moving velocity \cite{Xiong-2023}. While for a slowly moving incident soliton, the condition for the semi-classical theory is no longer valid and more sophisticated quantum theory need to be developed. 
In this work, we therefore develop a quantum scattering approach to address the problem of time-domain diffraction of matter-wave solitons.
We consider a one-dimensional BEC soliton interacting with a vibrating mirror created by an evanescent light. Under KH transformation, we then can exploit Dyson series and the quantum scattering theory to investigate the dynamics of the soliton that is reflected by the mirror. Our analytical theory can quantitatively deduce the locations and the relative weights of the scattered wave packets. Our predicted locations and weights of diffracted wave packets show a better agreement with numerical evolution's results than the semi-classical approach as well as the perturbative theory. In particular, when the atomic mirror is vibrating with the two-frequency form, the diffracted wave packets show multi-peak sideband structures in its momentum distribution.
Underlying mechanism has been uncovered by our scattering theory.

\section{physical model and theoretical formulation}

\subsection{Physical model}

\begin{figure}[htbp]
	\centering
	\includegraphics[width=85mm]{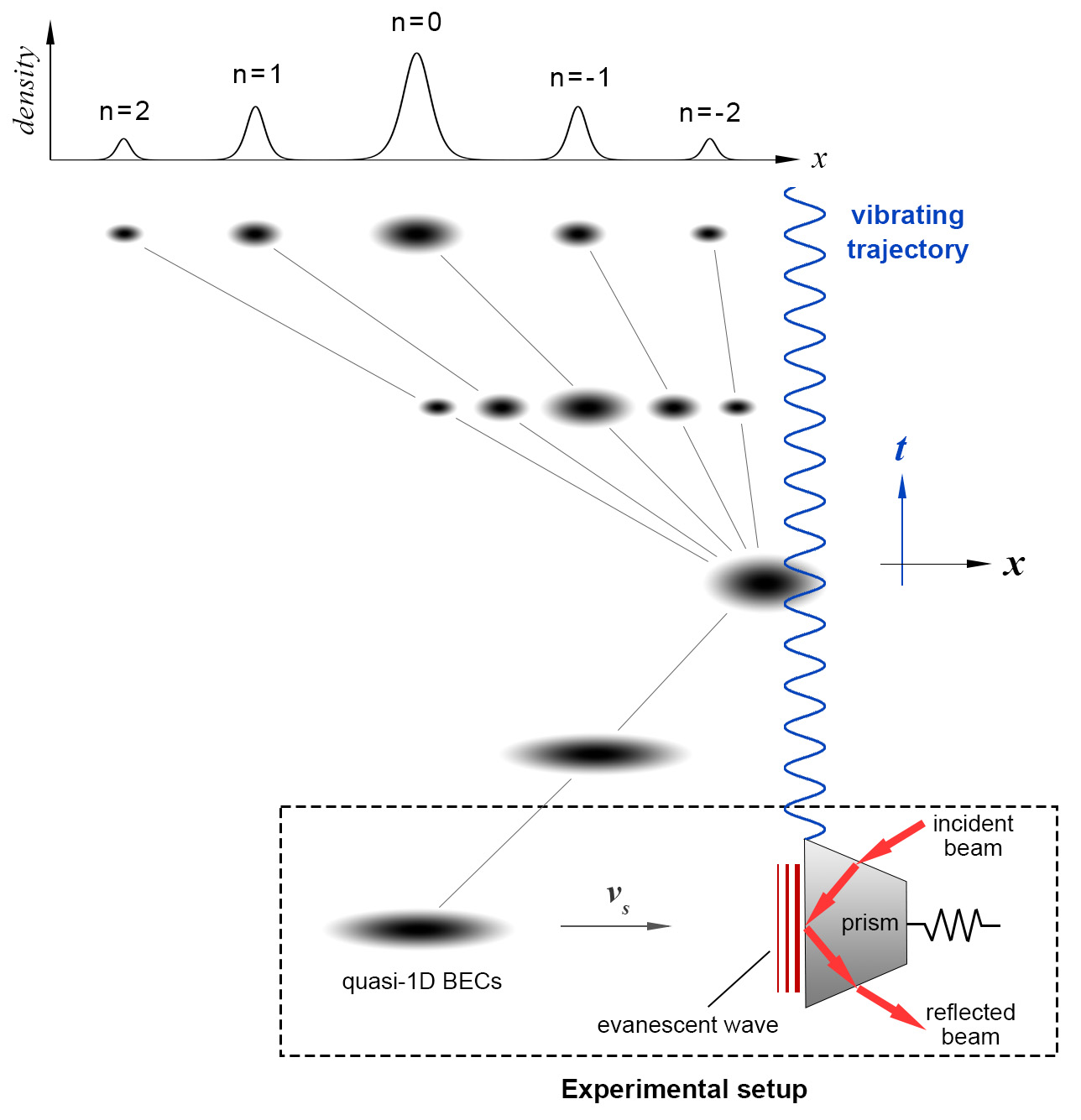}
	\caption{(Color online) Schematic diagram of the time-domain diffraction of quasi-1D BECs by a vibrating atomic mirror. 
    As shown in the box, the atomic mirror is made of a laser beam generating the evanescent wave (see the red arrows and lines) and a prism connecting with a spring (see the black broken line). The blue curve denotes the vibrating trajectory of mirror with time, and the grey arrow denotes the initial moving directions of soliton.
    The final density profile of BECs is shown at the top of the plot.
	}
	\label{pic-setup}
\end{figure}

We consider a physical process that an atomic wave of BECs interacts with a vibrating atomic mirror.
As shown in the box of Fig. \ref{pic-setup}, the quasi-1D BECs initially has a localized density distribution, moving towards the vibrating atomic mirror made from the glass prism shined by a laser beam.
When the laser beam is totally reflected by the inner surface of prism, the evanescent wave generating an exponentially decaying field appears outside the surface.
The frequency of the laser is blue-detuning with respect to the atomic levels in BECs.
As shown in Fig. \ref{pic-setup}, after reflected by the vibrating mirror, the matter-wave soliton can split into several soliton-like wave packets, showing the so-called time-domain diffraction phenomenon, in analogy with the space-domain diffraction of light waves by a reflection grating \cite{Hecht-book} or atomic waves by periodic potentials \cite{Estermann-1930,Moskowitz-1983,Gould-1986,Keith-1988,Carnal-1991}.

The dynamics of BECs can be described by the three-dimensional (3D) Gross-Pitaevskii model \cite{Kevrekidis-book},
\begin{align}\label{eq-3d}
	i\hbar\frac{\partial}{\partial {t}}\Psi(\mathbf {r},{t})=\Bigg[&-\frac{\hbar^{2}}{2m}\nabla^2+V_{\rm trap}(\mathbf {r})\nonumber\\
	&+V_{\rm mir}(x,{t})+g_{\rm 3D}|\Psi(\mathbf {r},{t})|^2\Bigg]\Psi(\mathbf {r},{t}),
\end{align}
where $m$ is the mass of atom.
$g_{\rm 3D}=4\pi\hbar^{2}a_{s}/{m}$ is the nonlinear coefficient, and $a_s$ is the $s$-wave scattering length.
The transverse trap potential has the form of $V_{\rm trap}(\mathbf {r})=m\omega_{\bot}^{2}({y}^{2} +{z}^{2})/2,$ where $\omega_{\bot}$ is the trap frequency.
The potential of atomic mirror is $V_{\rm mir}(x,{t})=V_{0}\,e^{2\kappa[{x}-{x}_{m}({t})]},$ where $V_0$ and $\kappa$ are respectively the potential's strength and decay factor. ${x}_{m}(t)$ is its time-dependent position and is set as a sine form here, $x_{m}(t)=a_{m}\sin(\omega_{m}t),$ where $a_{m}$ and $\omega_{m}$ are respectively the vibrating amplitude and frequency.

By the ansatz $\Psi(\mathbf {r},{t})=\mathscr{{\psi}}({x},{t}){\psi}_{\bot}({y},{z})e^{-i\omega_{\bot} {t}}$ (where ${\psi}_{\bot}({y},{z})={\rm exp}[-({{y}}^{2}+ {{z}}^{2})/2l_{\bot}^2]/(l_{\bot}\sqrt{\pi})$ and $l_{\bot}=\sqrt{\hbar/m\omega_{\bot}}$) and integrating the atom number's density on $y$ and $z$ directions, the 3D model (\ref{eq-3d}) can be transfered into the 1D model,
\begin{align}\label{eq-1d}
	i\hbar\frac{\partial}{\partial {t}}\mathscr{{\psi}}({{x}},{t})=\Bigg[-\frac{\hbar^{2}}{2m}\frac{\partial^2}{\partial
	{x}^2}+V_{\rm mir}({{x}},{t})+g_{\rm 1D} |{\psi}({{x}},{t})|^2\Bigg]{\psi}({{x}},{t}),
\end{align}
where $g_{1D}=2\hbar\omega_{\bot} a_{s}$ is the strength of one-dimensional nonlinearity.
When $a_s<0$ and $V_{\rm mir}({{x}},{t})=0$, Eq. (\ref{eq-1d}) supports the bright soliton solution \cite{Zakharov-1971},
\begin{align}\label{eq-bs}
	\psi_s(x,t)=\frac{l_\perp}{w_s\sqrt{2|a_{s}|}} \,{\rm sech}\Big[\frac{x-v_s\,t}{w_s}\Big]e^{i(k_sx-\mu t)},
\end{align}
where $v_s$ represents the soliton's velocity, $w_s$ determines its amplitude and width, $k_s=mv_s/\hbar$ is the wave number, and $\mu=\frac{\hbar}{2m}(k_s^2-1/w_s^2)$ is the chemical potential.
The number of atoms in soliton is $N_s=l_\perp^2/w_s|a_s|$.

\subsection{KH transformation and gauge potential}

As we know, KH transformation can provide a transition approach between a moving coordinate and a static one \cite{Kramers-1956,Henneberger-1968,Reed-1990,Wu-2022}, through which the time-dependent part of an external potential can be cast to a gauge potential.
The unitary KH transformation has the form of
\begin{align}
	&\psi'=\Omega\psi,\quad \Omega=\exp\Big\{\frac{i}{\hbar}\int_{0}^{t}[-\dot{x}_m(\tau)\hat{p}+\frac{1}{2}m\dot{x}_{m}^2(\tau)]d\tau\Big\}.
\end{align}
After it is applied into the model (\ref{eq-1d}), the model becomes
\begin{align}\label{eq-1dkh}
	i\hbar\frac{\partial}{\partial t}\psi'(x,t)=\Big[\frac{(\hat{p}-qA)^2}{2m}+V_{0}\,e^{2\kappa x}+g_{\rm 1D}|\psi'(x,t)|^2\Big]\psi'(x,t),
\end{align}
where $\hat{p}=-i\hbar\frac{\partial}{\partial x}$ is the momentum operator and $qA=m\dot{x}_{m}.$
$q$ and $A$ are respectively the effective particle's charge and the effective vector potential, and $\dot{x}_{m}={dx_{m}}/{dt}$ is the instantaneous velocity of mirror. 
It indicates that the vibration of mirror has been transformed into the gauge potential $A$, and the BECs interacts with a static mirror in the gauge potential.
It provides us a novel perspective to treat the time-domain diffraction phenomenon induced by a vibrating mirror.
The idea of artificial effective gauge potentials has been also presented to manipulate many kinds of microscopic particles, like neutral atoms \cite{Dalibard-2011} and photons \cite{Fang-2012,Rechtsman-2013}.

\subsection{Quantum scattering theory}

We can rewrite the model (\ref{eq-1dkh}) as
\begin{align}\label{eq-1dl}
		&i\hbar\frac{\partial}{\partial t}\psi'(x,t)=\Big[-\frac{\hbar^2}{2m}\frac{\partial^2}{\partial x^2}+V_{S}(x)+\hat{V}_{D}(t)\Big]\psi'(x,t),
\end{align}
where $V_{S}(x)=V_{0}\,e^{2\kappa x}$ and $\hat{V}_{D}(t)=i\hbar\dot{x}_{m}(t)\frac{\partial}{\partial x}+\frac{1}{2}m\dot{x}_{m}^2(t)+g_{\rm 1D}|\psi'(x,t)|^2$ are respectively the static and dynamical parts of potentials.
We define two Hamiltonians,
\begin{subequations}
	\begin{align}
		&\hat{H}_{SD}^{(t)}=-\frac{\hbar^2}{2m}\frac{\partial^2}{\partial x^2}+V_{S}{(x)}+\hat{V}_{D}{(t)},\\
		&\hat{H}_{D}^{(t)}=-\frac{\hbar^2}{2m}\frac{\partial^2}{\partial x^2}+\hat{V}_{D}{(t)},
	\end{align}
\end{subequations}
to describe the systems with and without the static field $V_{S}{(x)}$, respectively.
Their corresponding time evolution operators are
\begin{subequations}
	\begin{align}
		&\hat{U}_{SD}^{(t_2,t_0)}=\exp[-\frac{i}{\hbar}\int_{t_0}^{t_2}\hat{H}_{SD}^{(t_1)}dt_1],\\
		&\hat{U}_{D}^{(t_2,t_0)}=\exp[-\frac{i}{\hbar}\int_{t_0}^{t_2}\hat{H}_{D}^{(t_1)}dt_1].
	\end{align}
\end{subequations}
To apply the quantum scattering theory, as the first-order approximation, we ignore the nonlinear potential term of $g_{1D}|\psi_s|^2$.
Using the Dyson expansion \cite{Reiss-1992,Joachain-2012}, the two operators have the following relationship,
\begin{align}\label{eq-UD}
		&\hat{U}_{SD}^{(t_2,t_0)}=\hat{U}_{D}^{(t_2,t_0)}-\frac{i}{\hbar}
		\int_{t_0}^{t_2}\hat{U}_{D}^{(t_2,t_1)}V_{S}^{(x)}\hat{U}_{SD}^{(t_1,t_0)}	dt_1.
\end{align}
Then, one can apply Eq. (\ref{eq-UD}) to calculate the transition probability amplitude of atomic waves from an initial state $\psi_i$ into a final state $\psi_f$ (from time $t_0$ to $t_2$),
\begin{eqnarray}\label{eq-m-0}
		&M_{fi}^{(t_2,t_0)}=\langle\psi_f'^{(x,t_2)}|\hat{P}\,\hat{U}_{SD}^{(t_2,t_0)}|\psi_i'^{(x,t_0)}\rangle,
\end{eqnarray}
where $\hat{P}$ is the even parity operator produced by the reflection of wave packets, namely $\hat{P}\psi(x,t)=\psi(-x,t)$.
It is worth noting that the time-domain diffraction process happens in an effective gauge field and an evanescant light field, so it is always accompanied by the reflection of wave packets, which differs from the scattering process of particles in a realistic laser field \cite{Reiss-1992,Joachain-2012,Wu-2022-1,Liao-2022}.
Considering that the mirror is static in the new frame, one can conveniently use the parity operator $\hat{P}$ to take the contribution of reflection into account.
Now, the key issue is how to set the wave functions of the initial and final states.

Let us recall the physical process of the diffraction in the laboratory frame (namely the frame before KH transformation): a soliton-type wave packet moves towards a vibrating mirror and then interacts with it, and finally the wave packet is scattered into many discrete wave packets.
There are two main stages in the diffraction process: the initial stage before the diffraction phenomenon appears, and the final stage after that.
In the initial stage of diffraction, the wave packet is close with the mirror but is not oscillating with the mirror, so the wave function $\psi_i$ can be set as the eigenstate only under the static mirror's potential \cite{Henkel-1994},
\begin{eqnarray}
		&\psi_i^{(x,t)}=\frac{1}{\sqrt{L_i}}K_{\frac{ik_i}{\kappa}}\Big[\frac{\sqrt{2mV_0}}{\hbar\kappa}e^{\kappa x}\Big]e^{-i\omega_i t},
\end{eqnarray}
where $K_n[z]$ is the modified Bessel function of the second kind, and $L_i$ is a constant parameter with length unit.
For the initial state, its atom has the momentum $p_i=\hbar k_i=mv_s$ and the kinetic energy $E_i=\hbar\omega_i=p_i^2/2m$.
Thus, in the frame after KH transformation, the wave function of initial state is
\begin{align}\label{eq-psii}
	\psi_i'^{(x,t)}=\Omega\psi_i^{(x,t)}=\frac{1}{\sqrt{L_i}}K_{\frac{ik_i}{\kappa}}\Big[\frac{\sqrt{2mV_0}}{\hbar\kappa}e^{\kappa (x+x_{m}^{(t)})}\Big]e^{-i(\omega_i t+\varphi^{(t)})},
\end{align}
where the time-dependent phase $\varphi^{(t)}=\frac{1}{2\hbar}\int_{0}^{t}m\dot{x}_{m}^2(\tau)d\tau$ is produced by the second part of $\hat{V}_{D}(t)$.

In the final stage of diffraction, the scattered wave packets are far from the mirror, so the wave function $\psi_f$ can be set as plane waves,
\begin{eqnarray}
		&\psi_{f}^{(x,t)}=\frac{1}{\sqrt{L_f}}e^{ik_f x-i\omega_f t},
\end{eqnarray}
where $L_f$ is a constant parameter with length unit, and its atomic momentum and kinetic energy are respectively $p_f=\hbar k_f$ and $E_f=\hbar\omega_f=p_f^2/2m$.
Thus, in the frame after KH transformation, the wave function of final state is
\begin{eqnarray}\label{eq-fr}
		&\psi_{f}'^{(x,t)}=\Omega\psi_f^{(x,t)}=\frac{1}{\sqrt{L_f}}e^{ik_f(x+x_{m}^{(t)})-i(\omega_f t+\varphi^{(t)})},
\end{eqnarray}
which is the exact solution of Eq. (\ref{eq-1dl}) when the static potential $V_S$ is absent.
As we know, the motion of a free and charged particle in a plane-wave electromagnetic field can be exactly described by the Volkov state \cite{Reiss-1992,Volkov-1935}.
Therefore, considering the similarity between the effective field $\hat{V}_{D}^{(t)}$ and the electromagnetic field, the wave function (\ref{eq-fr}) can be regarded as the Volkov state in the effective field $\hat{V}_{D}^{(t)}$.

Then, one can apply Eqs. (\ref{eq-UD}) and (\ref{eq-m-0}) to calculate the transition probability amplitude,
\begin{align}\label{eq-m-1}
	M_{fi}^{(t_2,t_0)}&=\langle\psi_{f}'^{(x,t_2)}|\hat{P}\,\hat{U}_{SD}^{(t_2,t_0)}|\psi_i'^{(x,t_0)}\rangle\nonumber\\
	& =\langle\psi_{f}'^{(x,t_2)}|\hat{P}\,\hat{U}_{D}^{(t_2,t_0)}|\psi_i'^{(x,t_0)}\rangle\nonumber\\
	&\qquad -\frac{i}{\hbar}\int_{t_0}^{t_2}\langle\psi_{f}'^{(x,t_2)}|\hat{P}\,\hat{U}_{D}^{(t_2,t_1)}V_{S}^{(x)}\hat{U}_{SD}^{(t_1,t_0)}|\psi_i'^{(x,t_0)}\rangle dt_1\nonumber\\
	& \approx \langle\psi_{f}'^{(-x,t_0)}|\psi_i'^{(x,t_0)}\rangle
	-\frac{i}{\hbar}\int_{t_0}^{t_2}\langle\psi_{f}'^{(-x,t_1)}|\hat{V}_S^{(x)}|\psi_i'^{(x,t_1)}\rangle dt_1.
\end{align}
In Eq. (\ref{eq-m-1}), two approximations are used.
One of them is
\begin{eqnarray}\label{eq-appx1}
	&\langle\psi_{f}'^{(x,t_2)}|\hat{P}\,\hat{U}_{D}^{(t_2,t_1)}=\langle\psi_{f}'^{(-x,t_2)}|\hat{U}_{D}^{(t_2,t_1)}\approx \langle\psi_{f}'^{(-x,t_1)}|.
\end{eqnarray}
As $\hat{V}_D^{(t)}$ has the period $T=2\pi/\omega_m$, we can obtain $\hat{H}_D^{(t)}=\hat{P}\,\hat{H}_D^{(t+T/2)}$ and $\hat{U}_{D}^{(t_2,t_1)}=\hat{P}\,\hat{U}_{D}^{(t_2-T/2,t_1-T/2)}=\hat{P}\,\hat{U}_{D}^{(t_2-T/2,t_2)}\hat{U}_{D}^{(t_2,t_1)}\hat{U}_{D}^{(t_1,t_1-T/2)}$.
When $\hbar\omega_m$ is far larger than the atomic energy in the dynamical field, one can deduce that $\hat{U}_{D}^{(t_2-T/2,t_2)}$ and $\hat{U}_{D}^{(t_1,t_1-T/2)}$ approach to $1$ and obtain Eq. (\ref{eq-appx1}).
Accordingly, its approximative condition is $a_m\ll \hbar/mv_s$ and $\sqrt{\hbar/m\omega_m}$.
Another approximation is
\begin{eqnarray}
		&\hat{U}_{SD}^{(t_1,t_0)}|\psi_i'^{(x,t_0)}\rangle\approx |\psi_i'^{(x,t_1)}\rangle.
\end{eqnarray}
It is because that $\psi_i'^{(x,t)}$ can approximatively satisfy the model (\ref{eq-1dl}) when $V_{S}^{(x,t)}\approx V_0 e^{2\kappa(x+x_m^{(t)})}$ and $\hat{V}_{D}^{(t)}\approx i\hbar\dot{x}_{m}(t)\frac{\partial}{\partial x}+\frac{1}{2}m\dot{x}_{m}^2(t)$.
Accordingly, the approximative condition is $a_m\ll 1/\kappa$, namely a relative small vibrating amplitude is required.

Now, let us calculate the transition probability amplitude (\ref{eq-m-1}).
Its first part is
\begin{align}\label{eq-m1}
		M_{\rm I}^{(t_2)}&=\int_{-\infty}^{+\infty}\psi_f^{*(x,t_2)}\psi_i^{(x,t_2)}dx\nonumber\\
		&=\frac{1}{\sqrt{L_iL_f}}e^{i(\omega_f-\omega_i)t_2-ik_fx_{m}^{(t_2)}}\nonumber\\
		&\qquad\quad \times\int_{-\infty}^{+\infty}K_{\frac{ik_i}{\kappa}}\Big[\frac{\sqrt{2mV_0}}{\hbar\kappa}e^{\kappa (x+x_{m}^{(t_2)})}\Big]e^{ik_fx}dx\nonumber\\
		&=\frac{\phi_1}{\sqrt{L_iL_f}}\,e^{i(\omega_f-\omega_i)t_2-2ik_fx_{m}^{(t_2)}}\nonumber\\
		&=\frac{\phi_1}{\sqrt{L_iL_f}}\,\sum_{n=-\infty}^{+\infty}J_n(2a_mk_f)e^{i\Delta\omega t_2},
\end{align}
where $J_n[z]$ is the Bessel function of the first kind, and we define
\begin{align}
	\Delta \omega=\omega_f-\omega_i-n\omega_m.
\end{align}
Meanwhile, the relation $e^{i\alpha\sin \theta}=\sum_{n=-\infty}^{+\infty}J_n[\alpha]e^{in\theta}$ is used.
The function $\phi_1$ about $k_f$ is
\begin{align}
	\phi_1=\frac{1}{4\kappa}\Big(\frac{\sqrt{2}\hbar\kappa}{\sqrt{mV_0}}\Big)^{\frac{ik_f}{\kappa}}
	\Gamma\Big[\frac{i}{2\kappa}(k_f-k_i)\Big]	\Gamma\Big[\frac{i}{2\kappa}(k_f+k_i)\Big],
\end{align}
where $\Gamma[z]$ is the gamma function.

The second part of the transition probability amplitude (\ref{eq-m-1}) is
\begin{align}
	M_{\rm II}^{(t_0,t_2)}&=\frac{i}{\hbar}\int_{t_0}^{t_2}\int_{-\infty}^{+\infty}\psi_f^{*(x,t_1)}V_0 \,e^{2\kappa x}\psi_i^{(x,t_1)}dx\,dt_1\nonumber\\
	&=\frac{1}{\sqrt{L_iL_f}}\frac{i\,V_0}{\hbar}\int_{t_0}^{t_2}e^{i(\omega_f-\omega_i)t_1-ik_fx_m^{(t_1)}}\nonumber\\
	&\qquad\times\int_{-\infty}^{+\infty}K_{\frac{ik_i}{\kappa}}\Big[\frac{\sqrt{2mV_0}}{\hbar\kappa}e^{\kappa (x+x_{m}^{(t_1)})}\Big]e^{(ik_f+2\kappa)x}dx\,dt_1\nonumber\\
	&=\frac{\phi_2}{\sqrt{L_iL_f}}\frac{i\,V_0}{\hbar}\int_{t_0}^{t_2}e^{i(\omega_f-\omega_i)t_1-2(ik_f+\kappa)x_m^{(t_1)}}dt_1,
\end{align}
where the function $\phi_2$ about $k_f$ is
\begin{align}
	\phi_2=\frac{1}{4\kappa}\Big(\frac{\sqrt{2}\hbar\kappa}{\sqrt{mV_0}}\Big)^{\frac{ik_f}{\kappa}+2}
	\Gamma\Big[1+\frac{i}{2\kappa}(k_f-k_i)\Big]	\Gamma\Big[1+\frac{i}{2\kappa}(k_f+k_i)\Big].
\end{align}
Next, substituting $x_{m}(t)=a_{m}\sin(\omega_{m}t)$ into the above expression and using the relation $e^{i\alpha\sin \theta}=\sum_{n=-\infty}^{+\infty}J_n[\alpha]e^{in\theta}$, one can obtain
\begin{align}\
	M_{\rm II}^{(t_0,t_2)}&=\frac{\phi_2}{\sqrt{L_iL_f}}\frac{i\,V_0}{\hbar}\sum_{n=-\infty}^{+\infty}J_n[2a_m(k_f-i\kappa)]\int_{t_0}^{t_2}e^{i\Delta\omega t_1}dt_1,\nonumber\\
	&=\frac{\phi_2}{\sqrt{L_iL_f}}\frac{\,V_0}{\hbar}\sum_{n=-\infty}^{+\infty}J_n[2a_m(k_f-i\kappa)]\frac{e^{i\Delta\omega t_2}-e^{i\Delta\omega t_0}}{\Delta\omega}.\
\end{align}
Then, we consider $t_0=-t_2$,
\begin{align}
	M_{fi}^{(-t_2,t_2)}&=M_{\rm I}^{(t_2)}+M_{\rm II}^{(-t_2,t_2)}\nonumber\\
	&=\frac{1}{\sqrt{L_iL_f}}\,\sum_{n=-\infty}^{+\infty}\Big[\phi_1J_n(2a_mk_f)e^{i\Delta\omega t_2}\nonumber\\
	&\qquad + \frac{2i\phi_2V_0}{\hbar}J_n[2a_m(k_f-i\kappa)]\frac{\sin(\Delta\omega\, t_2)}{\Delta\omega}\Big].
\end{align}
When the interacting time is approaching infinity, the second part of $M_{fi}$ will be much larger than its first part (\ref{eq-m1}).
After neglecting its first part, its limit value can be written as
\begin{align}\label{eq-m-2}
	M&_{fi}^{(-\infty,+\infty)}=\lim_{t_2\rightarrow +\infty}M_{fi}^{(-t_2,t_2)}\nonumber\\
	&=\frac{1}{\sqrt{L_iL_f}}\frac{2i\phi_2V_0}{\hbar}\,\sum_{n=-\infty}^{+\infty}J_n[2a_m(k_f-i\kappa)]\lim_{t_2\rightarrow +\infty}\frac{\sin(\Delta\omega\, t_2)}{\Delta\omega}\nonumber\\
	&=\frac{1}{\sqrt{L_iL_f}}\frac{2i\pi\phi_2V_0}{\hbar\omega_\perp}\,\sum_{n=-\infty}^{+\infty}J_n[2a_m(k_f-i\kappa)]\delta\Big(\frac{\Delta\omega}{\omega_\perp}\Big),
\end{align}
where we use $\lim_{t\rightarrow+\infty}\frac{\sin\Omega t}{\Omega}=\pi\delta(\Omega).$

\begin{figure*}[htbp]
	\centering
	\includegraphics[width=175mm]{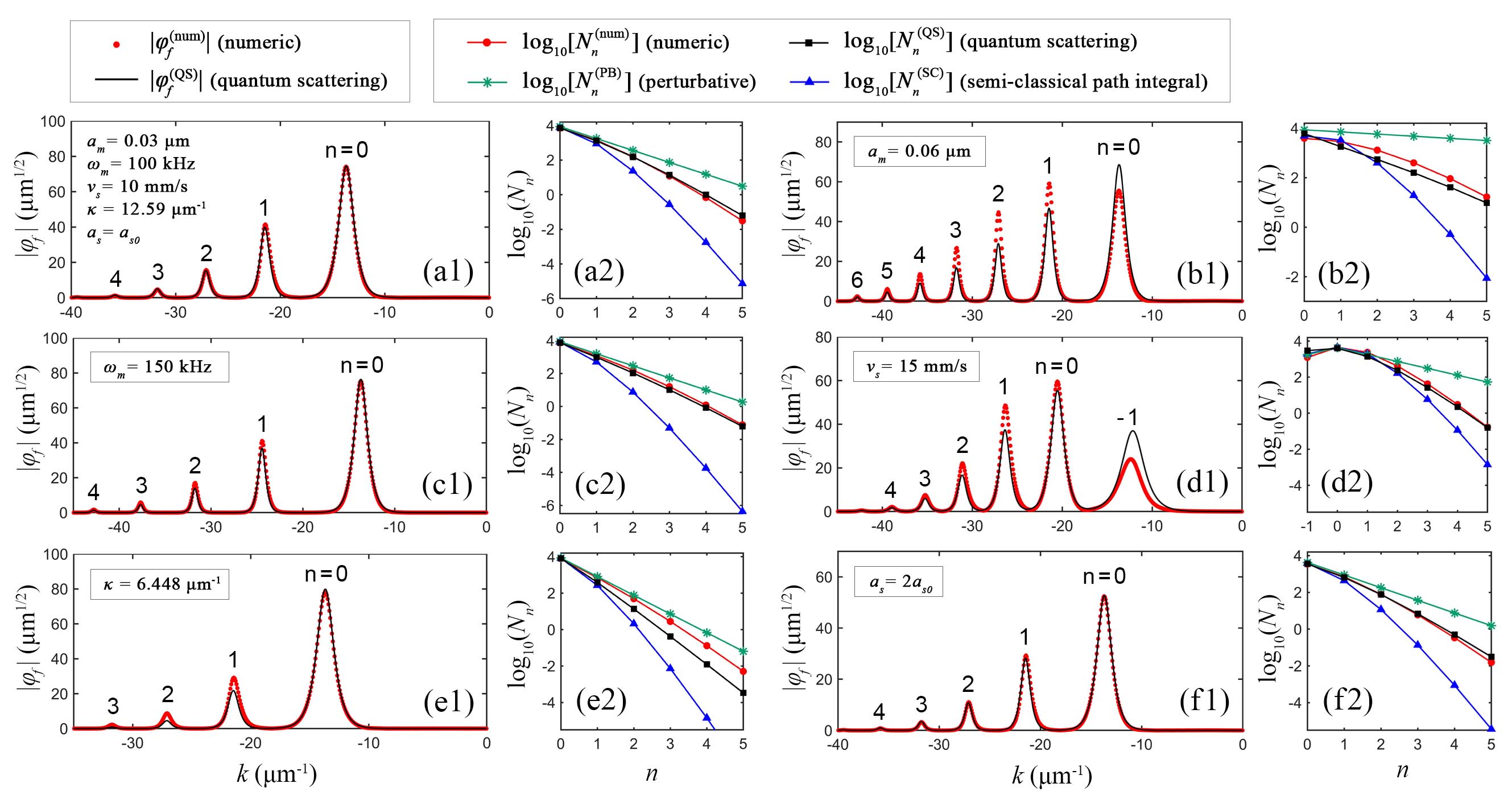}
	\caption{(Color online) (a1) Amplitude distribution of wave functions on momentum space and (a2) the atom number of the $n$-th wave packet, when $a_s=a_{s0}$, $\kappa=12.59\,{\rm \mu m^{-1}}$, $a_{m}=0.03\,{\rm \mu m}$, $\omega_{m}=100\,{\rm kHz}$, $w_s=1\,{\rm \mu m}$, $v_s=10\,{\rm mm/s}$. (b1-f2) Same as plot (a1,a2) except for (b1,b2) $a_{m}=0.06\,{\rm \mu m}$, (c1,c2) $\omega_{m}=150\,{\rm kHz}$, (d1,d2) $v_s=15\,{\rm mm/s}$, (e1,e2) $\kappa=6.448\,{\rm \mu m^{-1}}$, and (f1,f2) $a_s=2a_{s0}$. In the plots of amplitude distribution, the red dots and black curves are the results from numerical simulations and the quantum scattering theory, which correspond to Eq. (\ref{eq-phinum}) and Eq. (\ref{eq-phism}), respectively.
		In the plots of atom number, the red circles, black squares, green stars, and blue triangulars are the results from numerical simulations, the quantum scattering theory, the perturbative method, and the semi-classical path integral method, respectively.
	}
	\label{pic-of1}
\end{figure*}

From Eq. (\ref{eq-m-2}), one can see that $M_{fi}^{(-\infty,+\infty)}$ has the observable amplitude only when $\Delta\omega=0$. 
It means that the final state has discrete energy levels $E_{fn}=\hbar\omega_{fn}=\hbar(\omega_i+n\omega_m)$, and accordingly the final wave number is
\begin{align}\label{eq-kfn}
		&k_{fn}=-\sqrt{\frac{2m\omega_{fn}}{\hbar}}=-\sqrt{k_i^2+\frac{2m}{\hbar}n\omega_m}.
\end{align}
Thus, the transition probability amplitude from the initial state to the $n$-th final state is
\begin{align}
	M_n^{\rm (QS)}&=M_{fi}^{(-\infty,+\infty)}(k_f=k_{fn})\nonumber\\
	&=C_M\,\Gamma\Big[1+\frac{i}{2\kappa}(k_{fn}-k_i)\Big]	\Gamma\Big[1+\frac{i}{2\kappa}(k_{fn}+k_i)\Big]\nonumber\\
	&\quad\times J_n[2a_m(k_{fn}-i\kappa)],
\end{align}
where the superscript (QS) represents the results from the quantum scattering theory.
$C_M$ is a dimensionless coefficient with the form of
\begin{align}
	C_M=\frac{i\pi l_\perp^2\kappa}{\sqrt{L_iL_f}}\Big(\frac{\sqrt{2}\hbar\kappa}{\sqrt{mV_0}}\Big)^{\frac{ik_f}{\kappa}},
\end{align}
whose modulus is constant with respect to $k_f$.

\section{numerical results and discussion}

\subsection{Momentum distribution of atomic wave packets reflected by a one-frequency vibrating mirror}

Firstly, we numerically simulate the nonlinear Schr\"{o}dinger model (\ref{eq-1d}) by the split-step Fourier method \cite{Yang-book}.
The exact solution (\ref{eq-bs}) of soliton provides an initial condition for studying the collision between a soliton and an atomic mirror,
\begin{align}\label{eq-ini}
		\psi(x,t=0)=\frac{l_\perp}{w_s\sqrt{2|a_{s}|}} \,{\rm sech}\Big[\frac{x-x_0}{w_s}\Big]e^{imv_s(x-x_0)/\hbar},
\end{align}
where the initial velocity of soliton is controlled by $v_s$, and $x_0$ is a trivial quantity representing its initial position, which is set as different values so that the collision always happens when $t=2\,{\rm ms}$.
Also, the numerical evolution's wave function in the momentum space (i.e. the wavenumber space) can be calculated by the following Fourier transformation,
\begin{align}
	\varphi(k,t)=\frac{1}{\sqrt{2\pi}}\int_{-\infty}^{\infty} \psi(x,t)e^{-ikx}dx.
\end{align}
Its numerical array when $t=4\,{\rm ms}$ is used to compare with the analytical predictions, i.e.,
\begin{align}\label{eq-phinum}
	\varphi^{\rm (num)}_f(k)=\varphi(k,t=4\,{\rm ms}),
\end{align}
where the superscript (num) indicates a result from numerical simulation.

In this paper, we set the typical parameters as follows.
The mass of $^{87}{\rm Rb}$ atom is $m=1.445\times 10^{-25}\,{\rm kg}$.
According to Refs. \cite{Volz-2003,Becker-2008}, the $s$-wave scattering length is set as $a_s=a_{s0}$ or $2a_{s0}$, where the reference value of $s$-wave scattering length is $a_{s0}=-8.546\times 10^{-11}\,{\rm m}$.
The transversely trapping frequency is $\omega_\perp=2\pi\times 159\,{\rm Hz}$, so we have $l_\perp=0.855\,{\rm \mu m}$.
According to Ref. \cite{Voigt-2000}, one can set the evanescent wave's strength and decay factor as $V_0=1.807\times 10^{-28}\,{\rm J}$ and $\kappa=12.59\,{\rm \mu m^{-1}}$.
When $a_s=a_{s0}$, $\kappa=12.59\,{\rm \mu m^{-1}}$, $a_{m}=0.03\,{\rm \mu m}$, $\omega_{m}=100\,{\rm kHz}$, and $v_s=10\,{\rm mm/s}$, the numerical amplitude distribution of wave function in momentum space is shown in Fig. \ref{pic-of1} (a1).
Five peaks of different orders can be observed, which are the typical manifestation of a diffraction phenomenon.
By analyzing the momentum position of the peaks, we find that they are consistent with the result of Eq. (\ref{eq-kfn}) corresponding to the orders $n=0,1,2,3,4$.
To compare the numerical result with our analytic prediction, we find that all of the wave packets in Fig. \ref{pic-of1} (a1) approximatively has the shape of sech function, and their width decreases with $n$ increasing.
It inspires us to assume the wave function with the following expression,
\begin{align}\label{eq-phism}
	\varphi_f^{\rm (QS)}(k)=C_N\sum_{n=0}^{+\infty}M_n^{\rm (QS)}\sqrt{\frac{\pi w_s|k_{fn}|}{k_i}}{\rm sech}\Big[\frac{\pi w_s |k_{fn}|}{2k_i}(k-k_{fn})\Big],
\end{align}
where the wave packet of $n=0$ is assumed to have the same width as the initial one.
Meanwhile, before multiplying with $M^{\rm (QS)}_n$, the wave packet for every $n$ value has been normalized to ensure they have the same atom number. 
$C_N$ is a dimensionless coefficient to ensure that the total atom number equals to the initial one.
By comparison, our analytic prediction (see the black curve) has good agreement with the numerical one (see the red dots) for the peaks of all orders.
Then, we turn our attention into the atom numbers (or relative weights) of wave packets of different orders to quantitatively analyze the accuracy of our predictions.
We denote the atom number or the weight of the $n$-th wave packets by $N_n$.
In both of the numerical simulations and our quantum scattering method, it is calculated by
\begin{align}
	N^{\rm (num)}_n=\int_{k_{n-}}^{k_{n+}}|\varphi^{\rm (num)}_f(k)|^2dk,\;\; N^{\rm (QS)}_n=\int_{k_{n-}}^{k_{n+}}|\varphi^{\rm (QS)}_f(k)|^2dk,
\end{align}
where $k_{n\pm}=k_{f(n\pm 1/2)}$.
For a clearer observation, we use a semilog coordinate to show the atom number $N_n$ of the $n$-th wave packets. 
Fig. \ref{pic-of1} (a2) shows the results from the numerical simulation (red circles) and the quantum scattering method (black squares).
With $n$ increasing, the atom number decreases, and good agreements can be seen between the two results for wave packets of all orders.

The perturbative and semi-classical path integral methods have been also used to analyze the time-domain diffraction of an atomic wave \cite{Henkel-1994}.
Thus, it is interesting to compare the predictions from the two methods and our quantum scattering method.
For the perturbative method, according to Fermi’s golden rule, the transition probability amplitude is approximatively \cite{Henkel-1994,Dirac-1927,Fermi-1950}
\begin{align}
		M^{\rm (PB)}_{n} &\approx \big[M^{\rm (PB)}_1\big]^{|n|}\nonumber\\
		&=\Big[\frac{2\pi m\,a_m\omega_m}{\hbar\kappa}\frac{\sqrt{\sinh(\pi \, k_i/\kappa)\sinh(\pi \, |k_{f1}|/\kappa)}}{\cosh(\pi \, k_{f1}/\kappa)-\cosh(\pi \, k_i/\kappa)}\Big]^{|n|},	
\end{align}
where $M^{\rm (PB)}_1$ is the probability amplitude for the atom absorbing the energy $\hbar\omega_m$.
So the atom number of the $n$-th wave packet can be written as 
\begin{align}\label{eq-npb}
	N_n^{\rm (PB)}= N_s \,\big|M^{\rm (PB)}_n\big|^{2},	
\end{align}
where the superscript (PB) indicates a result from the perturbative method, and $N_s$ is the total atom number of BECs.
On the other hand, the transition probability amplitude from the semi-classical method can be written as \cite{Henkel-1994}
\begin{align}
M^{\rm (SC)}_n=J_n\Big[2a_mk_i\frac{\pi Q}{\sinh (\pi Q)}\Big]
\exp\Big({-inQ\ln\frac{V_0}{4\hbar\omega_i}}\Big),
\end{align}
where the superscript (SC) indicates a result from the semi-classical approach, and $Q={m\omega_m}/({2\hbar\kappa k_i})$.
Therefore, the atom number of the $n$-th wave packet can be calculated by
\begin{align}\label{eq-nsc}
	N_n^{\rm (SC)}= N_s \,\big|M^{\rm (SC)}_n\big|^{2}.
\end{align}
By Eqs. (\ref{eq-npb}) and (\ref{eq-nsc}), the atom numbers from the pertubative (green stars) and semi-classcal (blue triangulars) methods are shown in Fig. \ref{pic-of1} (a2).
For the perturbative method, $\log_{10}(N_n)$ is decreasing linearly with $n$ increasing, and its slope is equal to $2\log_{10}|M^{\rm (PB)}_1|$.
However, good agreements can be found only when $n$ is small, and the same conclusion is also obtained for the semi-classical method.

As the vibrating amplitude is increased to $a_{m}=0.06\,{\rm \mu m}$, the amplitude distribution of the wave function in momentum space is shown in Fig. \ref{pic-of1} (b1).
More peaks in the diffraction pattern appear, and our prediction from the quantum scattering theory shows a small deviation from the numerical result.
The deviation can be also seen in the corresponding distribution of atom number, i.e., Fig. \ref{pic-of1} (b2).
A larger vibrating amplitude $a_m$ also indicates a larger deviation between the perturbative method and the numerical result, as shown in Figs. \ref{pic-of1} (a2) and (b2).
Then, we also change some other parameters, such as the vibrating frequency $\omega_m$, the incident velocity $v_s$, the decay factor $\kappa$, and the $s$-wave scattering length $a_s$, and show the related results in Figs. \ref{pic-of1} (c1,c2), (d1,d2), (e1,e2), and (f1,f2), respectively.
Good agreements of our predictions with numerical results can be also observed in these cases, which indicates the effectiveness of the quantum scattering theory in these parameter ranges.
In particular, in Figs. \ref{pic-of1} (d1,d2), one can see the peak of the order $n=-1$ appears when the incident velocity is increased to $v_s=15\, {\rm mm/s}$.
Our detailed analysis suggests that the peak of the order $n=-1$ emerges when $v_s>\sqrt{2\hbar\omega_m/m}$.
Note that the semi-classical method is only applicable under the condition $k_i=mv_s/\hbar\gg \kappa$ provided by Ref. \cite{Henkel-1994}.

A recent work \cite{Xiong-2023} numerically studies the dynamics of a one-dimensional matter-wave soliton colliding with a vibrating atomic mirror.
They also find that the soliton splits into several wave packets with the discrete momentum corresponding to quantized kinetic energy after colliding.
Our quantum scattering approach can account for the main observations of Ref. \cite{Xiong-2023}, suppose a relative small vibrating amplitude, i.e., $a_m\ll \hbar/mv_s$, $\sqrt{\hbar/m\omega_m}$, and $1/\kappa$ as has been discussed in Sec. II C.

\subsection{Momentum distribution of atomic wave packets reflected by a two-frequency vibrating mirror}

As shown above, our quantum scattering approach has successfully applied to analyze the time-domain diffraction of matter waves by an atomic mirror with one-frequency vibration.
Here, in this section, we extend to apply our study to the case of two-frequency vibration.
The motion equation of atomic mirror with the two-frequency vibration can be written as
\begin{align}\label{eq-xm-2}
	&x_{m}(t)=a_{1}\sin(\omega_{1}t)+a_{2}\sin(\omega_{2}t),
\end{align}
where $a_{1}$, $a_{2}$ and $\omega_{1}$, $\omega_{2}$ are the amplitude and frequency of two vibrating modes, respectively.
Substituting the motion equation (\ref{eq-xm-2}) into the transition probability amplitude (\ref{eq-m-1}), one can obtain 
\begin{align}\label{eq-m-2-2}
	M&_{fi}^{(-\infty,+\infty)}=C_M\,\Gamma\Big[1+\frac{i}{2\kappa}(k_f-k_i)\Big]	\Gamma\Big[1+\frac{i}{2\kappa}(k_f+k_i)\Big]\nonumber\\
	&\quad\times\sum_{n=-\infty}^{+\infty}J_n[2a_1(k_f-i\kappa)]J_{n'}[2a_2(k_f-i\kappa)]\delta\Big(\frac{\Delta\omega'}{\omega_\perp}\Big),
\end{align}
where $\Delta \omega'=\omega_f-\omega_i-n\omega_{1}-n'\omega_{2}$.

Thus, the final state has discrete energy levels,
\begin{align}
    E_{fnn'}&=\hbar\omega_{fnn'}=\hbar\omega_i+n\hbar\omega_{1}+n'\hbar\omega_{2}.
\end{align}
Accordingly, the center wave number of the final wave packet of the order $(n,n')$ is
\begin{align}\label{eq-kfnn}
	k_{fnn'}&=-\sqrt{\frac{2m\omega_{fnn'}}{\hbar}}=-\sqrt{k_i^2+\frac{2m}{\hbar}(n\omega_{1}+n'\omega_{2})}.
\end{align}
The transition probability amplitude from the initial state to the final state of the order $(n,n')$ is
\begin{align}
	M_{nn'}^{\rm (QS)}&=C_M\,\Gamma\Big[1+\frac{i}{2\kappa}(k_{fnn'}-k_i)\Big]	\Gamma\Big[1+\frac{i}{2\kappa}(k_{fnn'}+k_i)\Big]\nonumber\\
	&\quad\times J_n[2a_1(k_{fnn'}-i\kappa)]\;
    J_{n'}[2a_2(k_{fnn'}-i\kappa)].
\end{align}
And the predicted wave function in momentum space can be written as
\begin{align}\label{eq-phiqs2}
	\varphi_f^{\rm (QS)}(k)=C_N\sum_{n,n'}&M_{nn'}^{\rm (QS)}\sqrt{\frac{\pi w_s|k_{fnn'}|}{k_i}}\nonumber\\
	&\times{\rm sech}\Big[\frac{\pi w_s|k_{fnn'}|}{2k_i}(k-k_{fnn'})\Big],
\end{align}
which will be used to compare with numerical results.

\begin{figure}[htbp]
	\centering
	\includegraphics[width=85mm]{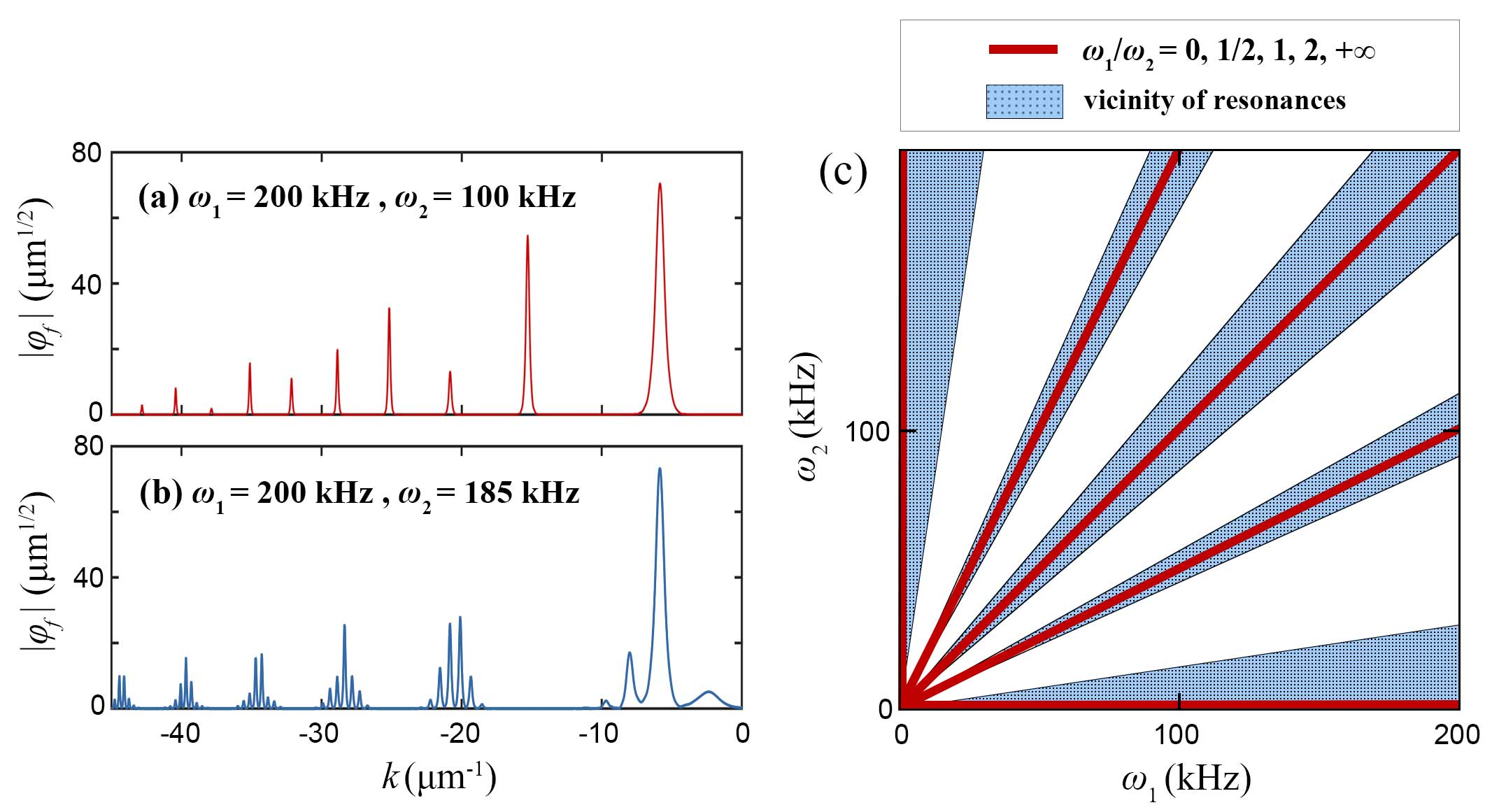}
	\caption{(Color online) (a-b) Typical distributions of diffracted wave packets: (a) $\omega_1=200\,{\rm kHz}$ and $\omega_2=100\,{\rm kHz}$ and (b) $\omega_1=200\,{\rm kHz}$ and $\omega_2=185\,{\rm kHz}$ from our numerical integration. (c) Phase diagram for the resonance regions of $\omega_1/\omega_2=0,1/2,1,2,+\infty$ (red solid lines).
	In the vicinity of the resonance (blue shadow regions), we find that the multi-peak sideband structures emerge.
	Other parameters are set as $a_s=a_{s0}$, $\kappa=12.59\,{\rm \mu m^{-1}}$, $a_{1}=a_{2}=0.05\,{\rm \mu m}$, $\omega_{1}=200\,{\rm kHz}$, $w_s=2\,{\rm \mu m}$, and $v_s=5\,{\rm mm/s}$. 
	}
	\label{pic-tf1}
\end{figure}

\begin{figure*}[htbp]
	\centering
	\includegraphics[width=176mm]{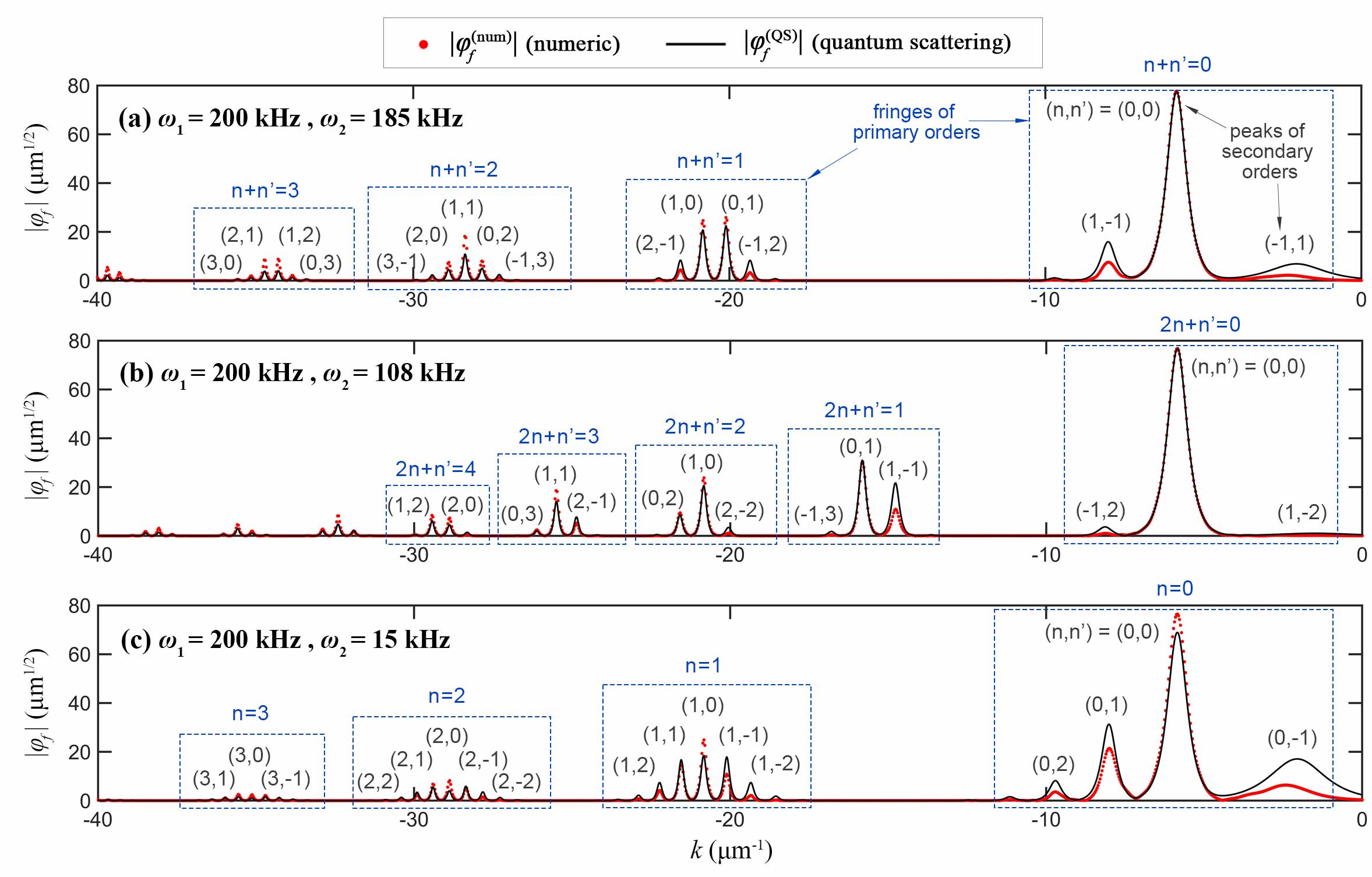}
	\caption{(Color online) Amplitude distribution of the wave function in momentum space when (a) $\omega_2=185\,{\rm kHz}$, (b) $\omega_2=108\,{\rm kHz}$, and (c) $\omega_2=15\,{\rm kHz}$. The red dots and black solid curves are respectively the results from numerical simulations and the quantum scattering theory. Other parameters are set as $a_s=a_{s0}$, $a_{1}=a_{2}=0.03\,{\rm \mu m}$, $\omega_{1}=200\,{\rm kHz}$, $w_s=2\,{\rm \mu m}$, and $v_s=5\,{\rm mm/s}$. 
	The fringes of primary orders are boxed by the blue rectangles, and the corresponding relations between $n$ and $n'$ are marked.
	The values of $(n,n')$ are marked on the peaks of secondary orders.
	}
	\label{pic-tf2}
\end{figure*}

Considering the two-frequency vibration and setting the typical parameters, $a_s=a_{s0}$, $a_{1}=a_{2}=0.05\,{\rm \mu m}$,  $w_s=2\,{\rm \mu m}$, and $v_s=5\,{\rm mm/s}$, $\omega_1=200\,{\rm kHz}$, and $\omega_2=100\,{\rm kHz}$, we numerically simulate the evolution of BECs and show the amplitude distribution of final state in Fig. \ref{pic-tf1} (a).
Some peaks of different orders can be observed.
However, different from the results of one-frequency vibration, the peak amplitude of these wave packets is not monotonically decreasing with $n$ increasing.
Its mechanism can be interpreted by the resonance of $\omega_1$ and $\omega_2$.
The two frequencies has the resonance relation $\omega_1/\omega_2=2$, so the amplitude of peaks is the superposition of the respective results of two frequencies, which may break the distribution rule of monotone decreasing.
Similar result has been also observed in Ref. \cite{Xiong-2023}, where the two frequencies have the resonance of $\omega_{1}/\omega_{2}=2/3$.

When we choose $\omega_1=200\,{\rm kHz}$ and $\omega_2=185\,{\rm kHz}$, an interesting phenomenon termed as multi-peak sidebands emerges, as shown in Fig. \ref{pic-tf1} (b).
It indicates that the multi-peak sideband structures may appear in the vicinity of the resonances.
Thus, we furthermore analyze the feature of diffraction patterns in the frequency range of $0<\omega_{1}<200\,{\rm kHz}$ and $0<\omega_{2}<200\,{\rm kHz}$.
As shown in Fig. \ref{pic-tf1} (c), we find that the multi-peak sideband structures emerge in the vicinity of resonances (blue shadow regions).

To figure out the formation mechanism of the interesting multi-peak sideband structure, we choose three groups of typical parameters near resonances, i.e., $\omega_2=185\,{\rm kHz}$, $108\, {\rm kHz}$, and $15\,{\rm kHz}$ with a fixed $\omega_{1}=200\,{\rm kHz}$.
The results are respectively shown in Figs. \ref{pic-tf2} (a), (b), and (c).
They show that our prediction of Eq. (\ref{eq-phiqs2}) always agrees well with the numerical results.
When $\omega_{2}=185\,{\rm kHz}$, as shown in Fig. \ref{pic-tf2} (a), the obvious multi-order diffraction pattern appears in the distribution of reflected wave packets.
Each primary fringe (labeled by the blue rectangles in Fig. \ref{pic-tf2}) contains a secondary-order multi-peak sideband structure.
After comparing Eq. (\ref{eq-kfnn}) and the position of peaks in the numerical distribution, the values of $n$ and $n'$ are marked.
For a better understanding, we can decompose the multi-photon energy absorption of the matter wave after reflected into the following form,
\begin{align}
		\Delta E =n\hbar\omega_{1}+n'\hbar\omega_{2}=(n+n')\hbar\frac{\omega_1+\omega_2}{2}+(n-n')\hbar\frac{\omega_1-\omega_2}{2}.
\end{align}
In the above equation, the sum frequency terms determine the fringes of primary orders and the difference frequency terms predict the multi-peak sideband structure of the secondary order.
Thus, the energy difference between adjacent fringes of primary orders is $\hbar\frac{\omega_1+\omega_2}{2}$, while the energy difference between adjacent peaks in each primary fringe is $\hbar\frac{\omega_1-\omega_2}{2}$.

When $\omega_2=108\,{\rm kHz}$, the multi-peak sideband structures can also emerge, as shown in Fig. \ref{pic-tf2} (b).
Similarly, we can decompose the multi-photon energy absorption into
\begin{align}
		\Delta E =(2n+n')\hbar\frac{\omega_1+2\omega_2}{2}+(2n-n')\hbar\frac{\omega_1-2\omega_2}{2}.
\end{align}
Thus, the energy differences between adjacent fringes of primary orders is $\hbar\frac{\omega_1+2\omega_2}{2}$, and the energy differences between adjacent peaks of secondary orders is $\hbar\frac{\omega_1-2\omega_2}{2}$.
Similarly, when $\omega_2=15\,{\rm kHz}$, the multi-peak sideband structures can be seen in Fig. \ref{pic-tf2} (c).
The two energy differences are just $\hbar\omega_1$ and $\hbar\omega_{2}$, due to $\omega_1\gg \omega_2$.

Here, we would like notice that the multi-peak sideband structures also appeared in the fluorescence spectrum of a two-level atom in a bichromatic optical field \cite{Ficek-1996,Tewari-1990,Zhu-1990,Freedhoff-1990}.
However, the underlying mechanism is quite different.
They originate from Rabi oscillation of the atom driven by the optical field.

\section{Conclusion}

Manipulating the motion of matter waves by the atomic mirror made from an evanescent wave is a research topic of great significance. We develop a non-perturbative quantum scattering theory to study the time-domain diffraction of matter-wave solitons interacting with a vibrating atomic mirror. Compared with the previous semi-classical or perturbative theory, our theory provides an alternative physical picture and shows a better agreement with numerical results. In particular, in the case of two-frequency vibration, our theory predicts the interesting multi-peak sideband structures in the diffraction patterns.
These theoretical predictions can be observable with current experimental techniques.
Otherwise, in our theoretical discussion, the nonlinear atomic interaction is ignored. Extending the quantum scattering theory to the nonlinear case is a challenging but might be important task and is worthy of future's study.
The related works are undergoing.

\section*{Acknowledgement}
The authors thank Prof. Di-Fa Ye and Dr. Binbing Wu for their helpful discussions.
This work was supported by NSAF (No.U2330401) and National Natural Science Foundation of China (No. 12247110).


\begin{thebibliography}{}

\bibitem{Dowling-1996} J. P. Dowling and J. Gea-Banacloche, Evanescent light-wave atom mirrors, resonators, waveguides, and traps, Adv. Atom. Mol. Opt. Phy. 37, 1 (1996).
\bibitem{Cook-1982} R. J. Cook, R. K. Hill, An electromagnetic mirror for neutral atoms, Opt. Commun. 43, 258 (1982).
\bibitem{Balykin-1998} V. I. Balykin, V. S. Letokhov, Y. B. Ovchinnickov, Quantum-state-selective mirror reflection of atoms by laser light, Phys. Rev. Lett. 60, 2137 (1998).
\bibitem{Henkel-1994} C. Henkel, A.M. Steane, R. Kaiser and J. Dalibard, A modulated mirror for atomic interferometry, J. Phys. II France 4, 1877 (1994).
\bibitem{Steane-1995} A. Steane, P. Szriftgiser, P. Desbiolles, and J. Dalibard, Phase Modulation of Atomic de Broglie Waves, Phys. Rev. Lett. 74, 4972 (1995).
\bibitem{Arndt-1995} M. Arndt, P. Szriftgiser, J. Dalibard, and A. Steane, Atom optics in the time domain, Phys. Rev. A 53, 3369 (1995).
\bibitem{Colombe-2005} Y. Colombe, B. Mercier, H. Perrin, and V. Lorent, Diffraction of a Bose-Einstein condensate in the time domain, Phys. Rev. A 72, 061601(R) (2005).

\bibitem{Estermann-1930} I. Estermann and O. Stern, Beugung von molekularstrahlen, Z. Phys. 61, 95 (1930).
\bibitem{Moskowitz-1983} P. E. Moskowitz, P. L. Gould, S. R. Atlas, and D. E. Pritchard, Diffraction of an atomic beam by standing-wave radiation, Phys. Rev. Lett. 51, 370 (1983).
\bibitem{Gould-1986} P. L. Gould, G. A. Ruff, and D. E. Pritchard, Diffraction of atoms by light: the near-resonant Kapitza-Dirac effect, Phys. Rev. Lett. 56, 827 (1986).
\bibitem{Keith-1988} D. W. Keith, M. L. Schattenburg, Henry I. Smith, and D. E. Pritchard, Diffraction of atoms by a transmission grating, Phys. Rev. Lett. 61, 1580 (1988).
\bibitem{Carnal-1991} O. Carnal and J. Mlynek, Young's Double-Slit Experiment with Atoms: A Simple Atom Interferometer, Phys. Rev. Lett. 66, 2689 (1991).


\bibitem{Chen-1996} W. Y. Chen, G. J. Milburn, and S. Dyrting, Effect of noise and modulation on the reflection of atoms from an evanescent wave, Phys. Rev. A 54, 1510 (1996).
\bibitem{Felber-1996} J. Felber, R. G\"{a}hler, and C. Rausch, Matter waves at a vibrating surface: Transition from quantum-mechanical to classical behavior, Phys. Rev. A 53, 319 (1996).
\bibitem{Xiong-2023} W. Xiong, P. Gao, Z. Y. Yang, and W. L. Yang, Quantized reflection of a soliton by a vibrating atomic mirror, Phys. Rev. A 108, 023303 (2023).
\bibitem{Kevrekidis-book} P. G. Kevrekidis, D. J. Frantzeskakis, and R. Carretero-Gonz$\acute{\rm a}$lez, Emergent Nonlinear Phenomena in Bose-Einstein Condensates: Theory and Experiment (Springer, New York, 2008).
\bibitem{Abdullaev-2005} F. Kh. Abdullaev, A. Gammal, A. M. Kamchatnov, and L. Tomio, Dynamics of bright matter wave solitons in a Bose–Einstein condensate, Int. J. Mod. Phys. B 19, 3415 (2005).
\bibitem{Frantzeskakis-2010} D. J. Frantzeskakis, Dark solitons in atomic Bose–Einstein condensates: from theory to experiments, J. Phys. A-Math. Theor. 43, 213001 (2010).
\bibitem{Kengne-2021} E. Kengne, W. M. Liu, and B. A. Malomed, Spatiotemporal engineering of matter-wave solitons in Bose–Einstein condensates, Phys. Rep. 899, 1 (2021).
\bibitem{Wu-2002} B. Wu, J. Liu, and Q. Niu, Controlled generation of dark solitons with phase imprinting, Phys. Rev. Lett. 88, 034101 (2002).
\bibitem{Zhao-2020} L. C. Zhao, X. W. Luo, and C. Zhang, Magnetic stripe soliton and localized stripe wave in spin-1 Bose-Einstein condensates, Phys. Rev. A 101, 023621 (2020).
\bibitem{Zhao-2020-2} L. C. Zhao, W. Wang, Q. Tang, Z. Y. Yang, W. L. Yang, and J. Liu, Spin soliton with a negative-positive mass transition, Phys. Rev. A 101, 043621 (2020).

\bibitem{Hecht-book} E. Hecht, Optics (Addison-Wesley, Reading, MA, 2001), Chap. 10. 2. 8.

\bibitem{Zakharov-1971} V. E. Zakharov and A. B. Shabat, Zh. Eksp. Teor. Fiz. 61, 118 (1971) [Sov. Phys. JETP 34, 62 (1971)].

\bibitem{Kramers-1956} H. A. Kramers, Collected Scientijic Papers (North-Holland,
Amsterdam, 1956), p. 272.
\bibitem{Henneberger-1968} W. C. Henneberger, Perturbation method for atoms in intense light beams, Phys. Rev. Lett. 21, 838 (1968).
\bibitem{Reed-1990} V. C. Reed and K. Burnett, Ionization of atoms in intense laser pulses using the Kramers-Henneberger transformation, Phys. Rev. A 42, 3152 (1990).
\bibitem{Wu-2022} B. Wu, H. Duan, and J. Liu, Resonant tunneling of deuteron-triton fusion in strong high-frequency electromagnetic fields, Phys. Rev. C 105, 064615 (2022). 

\bibitem{Dalibard-2011} J. Dalibard, F. Gerbier, G. Juzeli$\bar{u}$nas, and Patrik \"{O}hberg, Colloquium: Artificial gauge potentials for neutral atoms, Rev. Mod. Phys. 83, 1523 (2011).
\bibitem{Fang-2012} K. Fang, Z. Yu, and S. Fan, Realizing effective magnetic field for photons by controlling the phase of dynamic modulation, Nat. Photonics 6, 782 (2012).
\bibitem{Rechtsman-2013} M. C. Rechtsman, J. M. Zeuner, Y. Plotnik, Y. Lumer, D. Podolsky, F. Dreisow, S. Nolte, M. Segev, and A. Szameit, Photonic Floquet topological insulators, Nature 496, 196 (2013).

\bibitem{Reiss-1992} H. R. Reiss, Theoretical methods in quantum optics: S-matrix and Keldysh techniques for strong-field processes, Prog. Quant. Elec. 16, 1 (1992).
\bibitem{Joachain-2012} C. J. Joachain, N. J. Kylstra, and R. M. Potvliege, Atoms in intense laser fields (Cambridge University Press, Cambridge, 2012).
\bibitem{Wu-2022-1} B. Wu and J. Liu, Proton emission from halo nuclei induced by intense x-ray lasers, Phys. Rev. C 106, 064610 (2022).
\bibitem{Liao-2022} L. G. Liao, Q. Z. Xia, J. Cai, and J. Liu, Semiclassical trajectory perspective of glory rescattering in strong-field photoelectron holography, Phys. Rev. A 105, 053115 (2022).
\bibitem{Volkov-1935} D. M. Volkov, \"{U}ber eine klasse von l\"{o}sungen der diracschen gleichung, Z. Phys. 94, 250 (1935).

\bibitem{Yang-book} J. Yang, Nonlinear Waves in Integrable and Nonintegrable
Systems (Siam, Philadelphia, 2010).

\bibitem{Volz-2003} T. Volz, S. D\"{u}rr, S. Ernst, A. Marte, and G. Rempe, Characterization of elastic scattering near a Feshbach resonance in $^{87}$Rb, Phys. Rev. A 68, 010702(R) (2003).
\bibitem{Becker-2008} C. Becker, S. Stellmer, P. Soltan-Panahi, S. D\"{o}rscher, M. Baumert, E. M. Richter, J. Kronj\"{a}ger, K. Bongs, and K. Sengstock, Nat. Phys. 4, 496 (2008).
\bibitem{Voigt-2000} D. Voigt, B. T. Wolschrijn, R. Jansen, N. Bhattacharya, R. J. C. Spreeuw, and H. B. van Linden van den Heuvell, Observation of radiation pressure exerted by evanescent waves, Phys. Rev. A  61, 063412 (2000).

\bibitem{Dirac-1927} P. A. M. Dirac, The quantum theory of the emission and absorption of radiation, Proc. R. Soc. Lond. A 114, 243 (1927).
\bibitem{Fermi-1950} E. Fermi, Nuclear Physics (Chicago, IL: University of Chicago Press, 1950), p. 142.

\bibitem{Freedhoff-1990} H. Freedhoff and Z. Chen, Resonance fluorescence of a two-level atom in a strong bichromatic field, Phys. Rev. A 41, 6013 (1990).
\bibitem{Tewari-1990} S. P. Tewari and M. K. Kumari, Spectrum of resonance fluorescence from a two-level atom interacting with 100\% amplitude-modulated intense radiation, Phys. Rev. A 41, 5273 (1990).
\bibitem{Zhu-1990} Y. Zhu, Q. Wu, A. Lezama, D. J. Gauthier, and T. W. Mossberg, Resonance fluorescence of tvvo-level atoms under strong bichromatic excitation, Phys. Rev. A 41, 6574 (1990).
\bibitem{Ficek-1996} Z. Ficek and H. S. Freedhoff, Fluorescence and absorption by a two-level atom in a bichromatic field with one strong and one weak component, Phys. Rev. A 53, 4275 (1996).

\end{thebibliography}
\end{document}